\documentclass[twocolumn,superscriptaddress,amsmath,amssymb,pre]{revtex4}
\usepackage{graphicx}
\usepackage{dcolumn}
\usepackage{bm}
\usepackage{epsfig}
\usepackage[dvips]{color}
\usepackage{amssymb}
 
\begin{document}
\title{Three-dimensional aspects of fluid flows in channels. II. Effects of Meniscus and Thin Film regimes on Viscous Fingers}

\author{\firstname{R.} \surname{Ledesma-Aguilar}}
\email{rodrigo@ecm.ub.es}
\affiliation{Departament d'Estructura i Constituents de la Mat\`eria. Universitat de Barcelona, Avinguda Diagonal 647, E-08028 Barcelona, Spain}
\author{\firstname{I.} \surname{Pagonabarraga}}
\affiliation{Departament de F\'isica Fonamental.  Universitat de
  Barcelona, Avinguda Diagonal 647, E-08028 Barcelona, Spain}
\author{\firstname{A.} \surname{Hern\'andez-Machado}} 
\affiliation{Departament d'Estructura i Constituents de la Mat\`eria. Universitat de Barcelona, Avinguda Diagonal 647, E-08028 Barcelona, Spain}
\date{\today}

\begin{abstract}
We perform a three-dimensional study of steady state viscous fingers that develop in linear channels.  
By means of a three-dimensional Lattice-Boltzmann scheme that mimics the full macroscopic equations of
motion of the fluid momentum and order parameter, we study the effect of the thickness of the channel in two
cases.  
First, for total displacement of the fluids in the channel thickness direction, we find that the steady state
finger is effectively two-dimensional and that previous two-dimensional results can be recovered by taking into
account the effect of a curved meniscus across the channel thickness as a contribution to surface stresses. 
Secondly, when a thin film develops in the channel thickness direction, the finger narrows with increasing
channel aspect ratio in agreement with experimental results.  The effect of the thin film renders the problem 
three-dimensional and results deviate from the two-dimensional prediction.  

\end{abstract}
\maketitle
\section{Introduction}
Interfacial instabilities in three-dimensional channels give rise to a rich phenomenology in systems 
that range from nano and microscales\cite{Tabeling01} to macrometric channels\cite{deGennes02,couder,pelce}, and 
from which a number of practical applications can be drawn.

For instance, controlled drop breakup in micro-channels has proved useful in the fabrication of low polydispersity micro-emulsions\cite{Weitz01} and in the enhancement of micro-reaction processes\cite{Hosokawa01,Song01}.  
In the latter, three-dimensional effects are crucial, as they are responsible of a vortex flow structure within the 
droplet\cite{Kinoshita01} that enhances the mixing process of the reactants.  

A widely studied interfacial instability in channels is that of fingering, which occurs whenever a 
low-viscosity (or high-density) fluid drives a high-viscosity(or low-density) one.  The instability, 
first studied by Saffman and Taylor\cite{saffman}, leads to interface dynamics where finger-like structures 
emerge and compete.  The problem has a steady-state solution, composed by a single finger of constant velocity 
$U$ and occupies a fraction $\lambda$ of the width of the channel.  

Experimentally, finger growth has been studied mainly in Hele-Shaw cells. These consist of a pair of plates 
of length $L$ and width $W$ separated by a thickness $b$.  For such systems, it has been pointed out\cite{aref} 
that the stationary finger is determined by a single control parameter, a modified capillary number defined as $1/B=12Ca/\epsilon^2$.  
For a fluid with viscosity $\eta$ and surface tension $\sigma$, the capillary number, $Ca=\eta U/\sigma$, measures the 
competition between driving forces, such as viscous stresses and gravity, and restoring forces, like surface tension.   
$1/B$ also includes the degree of asymmetry of the cell, given 
by the aspect ratio $\epsilon=b/W$. If $1/B$ is the only control parameter of the system, all experimental data, 
\emph{i.e.} all finger widths, should be described by a single curve when plotted as a function of this parameter. 
Contrary to this view, experiments show that there exists a family of curves $\lambda$ \emph{vs.} $1/B$ for different aspect 
ratios\cite{libchaber,libchaber02}.  This fact suggests that a three-dimensional effect, given by the  
interplay between the dynamics in the channel-thickness and in the channel-width, is determinant for the 
steady-state solution. 

Theoretically, fluid-flow in a channel at small velocities pertains to the lubrication 
regime, in which the flow occurs mainly along the direction of $L$ given that it is much 
larger than both $W$ and $b$. Hele-Shaw flows are a limiting case in lubrication theory, where 
$b$ is much smaller than $W$.   Owing to the smallness of $b$, the problem is rendered effectively two-dimensional by averaging all fields over the thickness of the channel. Averaging 
the equations of motion also reduces the interface from a surface to a line, often called the \emph{leading interface}.  
In views of the averaged model, three-dimensional effects enter as perturbative corrections 
to the boundary conditions that hold at the leading interface in terms of $Ca$ and $\epsilon$, 
particularly to the Gibbs-Thomson condition, which relates the pressure drop across the interface to the interface curvature 
and surface tension. 

Progress towards a three-dimensional description of the problem has been made since the pioneering work of Saffman 
and Taylor\cite{saffman}, who solved the problem of a stationary finger in the absence of surface tension in two 
dimensions.  McLean and Saffman\cite{mclean} included the effect of surface tension and were the first to obtain 
a $\lambda$ \emph{vs.} $1/B$ prediction by solving numerically the two-dimensional model. According to their results, 
$\lambda$ is a monotonically decreasing function of $1/B$ that saturates to $\lambda \rightarrow 1/2$ as 
$1/B\rightarrow \infty$. The prediction of McLean and Saffman is unique in $1/B$, so the role of the aspect 
ratio is precluded from their theory.

The relevance of three-dimensional effects was first suggested by Park and Homsy\cite{parkhomsy}, who 
pointed out that a thin film of fluid in the channel-thickness direction would contribute to the pressure 
drop at the leading interface.  Using perturbation methods for slightly curved leading 
interfaces(small $\epsilon$), they found that for low $Ca$ the pressure drop varies as $Ca^{2/3}$, a result that 
matched the early prediction of Bretherton\cite{Bretherton01} for capillary tubes.

Sarkar and Jasnow\cite{jasnow} used the modified pressure drop to solve the steady 
state finger.  Their results agreed better with experiments but were restricted to 
low values of $Ca$. It was shown by Tabeling and Libchaber\cite{libchaber} 
that corrections to the pressure drop can be used to reduce three-dimensional experimental data
to the two-dimensional results of McLean and Saffman.  A modified pressure drop 
can be accounted for as an effective surface tension.  Using the correction of 
Park and Homsy, Tabeling and Libchaber were able to reduce their data to McLean and Saffman results 
for moderately low values of $1/B$, where fitting parameters were used to estimate the 
correction terms.  In an experimental study\cite{libchaber02}, Tabeling, Zocchi and Libchaber observed that, 
contrary to the McLean-Saffman prediction, the finger width can go below the one-half limit for sufficiently 
high $1/B$ and sufficiently large $\epsilon$. 

Reinelt extended the expansion of the pressure drop up to $O(1)$ in $Ca$ and included the effect of larger 
aspect ratios\cite{reinelt}.  Computation of the steady state finger yielded solutions that better agreed with 
experiments for $O(1)$ values of $Ca$.  For small $\epsilon$ Reinelt observed a better agreement 
between numerics and experiments.  However, for relatively large $\epsilon$ this agreement is lost.  

Higher $Ca$ values have only been explored in the case of flat leading interfaces by Halpern 
and Gaver\cite{Halpern01}. Their numerical results are consistent with results found by Reinelt and Saffman\cite{Reinelt02} 
for $Ca$ $O(1)$  and $\epsilon=0$, and show that the pressure drop is insensitive to the capillary number for $Ca > 20$.  

As an alternative to the sharp interface model, a number of mesoscopic approaches have gained 
importance in interface dynamics.  These are based on order parameter evolution equations of the 
Cahn-Hilliard type.  Being mesoscopic in nature, fluids are separated by diffuse 
regions instead of sharp interfaces, where the interface boundary conditions arise naturally.  
All mesoscopic models that address the viscous fingering problem so far are two-dimensional. 
For fluids of arbitrary viscosities and densities, Folch \emph{et al}\cite{folch,folch02} 
used a set of coupled evolution equations for the velocity potential and order parameter that 
describes accurately the early stages of destabilization of the leading interface, and  
approaches McLean and Saffman results as the viscosity of the displacing fluid is made negligible. 
The strict one-sided situation, were one of the fluids is inviscid, was studied by 
Hern\'andez-Machado~\emph{et al}\cite{aurora}. They used a single evolution equation for the concentration 
that includes dynamic effects in the form of chemical potential gradients and described correctly the steady state 
finger.  
   
In a preceding paper\cite{Ledesma01}, we have shown that a detailed three-dimensional description of fluid-flow in a channel
can be done by means of a mesoscopic model which we implement numerically \textit{via} a Lattice-Boltzmann 
algorithm.   The model considers a fluid-fluid interface in contact with solid boundaries.   In contrast to classic approaches, 
it allows for slip at the contact line by means of a diffusive mechanism inherent to the mesoscopic nature of the interface. 
This circumvents the complications of contact line dynamics in the classic formulation, associated to the  viscous dissipation singularity\cite{deGennes01}. 
In Ref.\cite{Ledesma01}, we focused on the case of a flat leading 
interface. We showed that depending on the velocity of the contact lines, which we control by modifying the diffusion strength, 
the interface can either advance as a meniscus or develop as a finger.  In the latter case, a thin film of displaced fluid is left adhered to the walls of the channel.

In this paper we extend our Lattice-Boltzmann simulations to the case of a non-flat leading 
interface, where a viscous finger is expected to appear.   
Our aim is to provide a detailed description of the mechanisms that affect the 
steady finger and that cause deviations from two-dimensional results. To do so, we study fingers that form 
in the meniscus and thin film regimes separately.  We cover values of $Ca$ up to $O(10)$ and explore various 
aspect ratios. 

The paper is organized in the following manner.  In Sec.~\ref{sec:HS} we present the governing equations of the system which we solve 
numerically by means of a Lattice Boltzmann algorithm, presented in the preceding paper\cite{Ledesma01}.
Results are presented in Sec.~\ref{sec:Results}. In 
Sec.~\ref{subsec:Param} we describe the simulation strategy and parameter steering procedure. As a validation test, in Sec.~\ref{subsec:LAS} we compute the dispersion relation of the interface in the two-dimensional limit and compare it to the analytic prediction of the Saffman-Taylor problem.  
Sec.~\ref{subsec:VF} is devoted to the study of stationary viscous fingers;  in Sec~\ref{subsubsec:VFM} we focus on fingers 
pertaining to the meniscus regime in the channel thickness, which we found to be effectively two-dimensional, 
while in Sec.~\ref{subsubsec:VFTF} fingers in the thin film regime are studied.  We find that fingers in the thin-film regime are three-dimensional and cannot be described by the two-dimensional theory in general.  A discussion of our results where we compare with previous experiments is presented 
in Sec.\ref{sec:Disc}. In Sec.~\ref{sec:Conclusions} we present the conclusions of this work. 

\section{Governing Equations}
\label{sec:HS}
\begin{figure}
\includegraphics[width=0.45\textwidth]{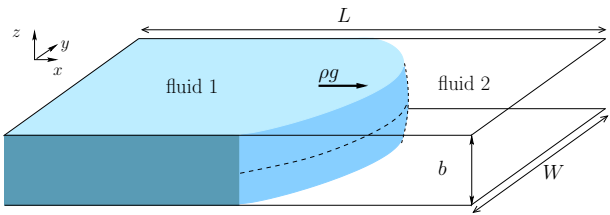}
\caption{Schematic representation of the system. \label{fig:geometry}  Dashed lines indicate 
projections of the fluid-fluid interface in the $xy$ and $xz$ planes. The leading interface corresponds to the 
$xy$ projection.}
\end{figure}
We consider the motion of two viscous fluids, whose dynamics are governed by the Navier-Stokes equations, 
\begin{equation}
\rho \left(\partial_t\vec v + \vec v (\vec \nabla \cdot \vec v)\right)  = - \vec \nabla P -\phi\vec \nabla \mu + 
\eta \nabla^2 \vec v + \rho \vec g. 
\label{eq:ns1}
\end{equation} 
Here $\vec v$ is the fluid velocity, $P$ is the pressure, $\rho$ is the density, $\eta$ is the fluid viscosity and $\vec g$ 
is the acceleration of gravity.  The extra term, $\phi \vec \nabla \mu$, is mesoscopic and accounts for interfacial forces. Here, $\phi(\vec r,t)$ is 
an order parameter and $\mu(\phi)$ is the chemical potential.  $\phi$ has the property of being uniform in the volume of 
each phase and non-uniform in an interfacial region of typical size $\xi$.  In the present case, volume values are chosen as 
$\phi= \pm 1$ for the displacing and the displaced fluid respectively, so the interface is located at $\phi=0$.  The size of the 
interface is set to $\xi=0.57$.

The dynamics of $\phi$ obey a convection-diffusion equation, 
\begin{equation}
\partial_t\phi + \vec v \cdot \vec \nabla \phi =  M \nabla^2\mu,
\label{eq:cd1}
\end{equation}
where $ M$ is a mobility coefficient. In equilibrium, the pressure and chemical potential 
minimize a free energy functional, from which explicit expressions $P(\rho)$ and $\mu(\phi)$
can be derived.  For further details the reader is referred to Ref.\cite{Ledesma01}.  

We work on a linear channel, composed by two solid plates of width $W$ and length $L$ parallel to the $xy$ plane,  
separated by a distance $b$, as depicted in Fig.~\ref{fig:geometry}.  There exist two principal directions in the system:  
a lateral direction, parallel to the $xy$ plane, and a transverse one parallel to the $xz$ plane. We will denote these by subscripts 
$\parallel$ and $\perp$ respectively.  

The impenetrability and stick boundary conditions at the walls are enforced by setting 
$\vec v (x,y,z=0) = \vec v (x,y,z=b) = 0 $ and  
$\phi \vec v(x,y,z=0) = \phi \vec v(x,y,z=b)= 0$. 
At both ends of the channel in the $x$ direction the flow is homogeneous. 
Hence, $\partial_x \rho  \vec v (x=0,y,z) = \partial_x \rho  \vec v (x=L,y,z) = 0$, and $ \partial_x \phi \vec v (x=0,y,z) =  
\partial_x \phi \vec v (x=L,y,z) = 0$.  Periodic boundary conditions are imposed in the $y$ direction.   

As for the fluid-fluid boundary, the Gibbs-Thomson relation is recovered by 
integrating Eq.(\ref{eq:ns1}) across the interfacial region, 
\begin{equation}
\Delta P = \sigma\left( \frac{1}{R_\parallel} + \frac{1}{R_\perp} \right), 
\label{gt}
\end{equation}    
where $\sigma$ is the surface tension and $R_\alpha$ is the radius of curvature of the interface 
in the direction $\alpha$.  

We now briefly review the classic treatment of the problem. First,  
$v_\perp$ is assumed to be much smaller than $v_\parallel,$  which in turn is expected to be parabolic in $z$.  
As a result, Eq.(\ref{eq:ns1}) is recast in the form of 
an average velocity field which holds in the volume of each fluid, called Darcy's Law, 
\begin{equation}
\langle v_\parallel \rangle  = -\frac{b^2}{12 \eta}\left(\vec \nabla P - \rho \vec g \right)_\parallel,
\label{eq:darcy}
\end{equation}
where triangular brackets denote an average over the channel thickness.  Under these conditions, $R_\perp$ 
is expected to be much larger than $R_\parallel$.  Hence, in the two-dimensional theory the Gibbs-Thomson 
relation is simplified to $\Delta P = \sigma/R_\parallel$.  

Corrections to this expression arise whenever $1/R_\perp$ is not negligible.  For such cases, Libchaber and Tabeling\cite{libchaber}
have proposed that thin film effects can be accounted for by defining  an effective surface tension
$$\sigma^* = \sigma \left(1+\frac{R_\parallel}{R_\perp}\right),$$
so the two-dimensional form of the Gibbs-Thomson condition is recovered.    For this purpose, 
they used the estimation of Park and  Homsy\cite{parkhomsy} of the pressure drop for  $Ca\rightarrow 0 $ and slightly curved leading interfaces($\epsilon \rightarrow 0$), 
\begin{equation}
\Delta P = \sigma\left(\frac{\pi}{4R_\parallel}+\frac{3.80}{b/2}Ca^{2/3}\right).
\label{eq:ph}
\end{equation}
As a result, their experimental results collapsed to the McLean-Saffman curve when using the corresponding 
definition of the control parameter, $1/B^*=(\sigma/\sigma^*)1/B$.  

We solve numerically Eqs.~(\ref{eq:ns1}) and (\ref{eq:cd1}) by means of a Lattice-Boltzmann algorithm.  
For further details of the method, the reader is referred to the preceding paper\cite{Ledesma01}.

\section{Results}
\label{sec:Results}
\subsection{Simulation Parameters and Setup}
\label{subsec:Param}
The traditional description of the viscous fingering problem corresponds to situations
in which the relevant forces at play are viscous stresses and capillarity.  For the particular 
case of fingering in a Hele-Shaw cell these forces are expressed in terms of a modified capillary 
number\cite{aref} $1/B=12(W/b)^2(\Delta \eta U +\Delta \rho g b^2/12)/\sigma,$ where $\Delta \eta$ and $\Delta 
\rho $ are the differences in viscosity and density between the fluids. 

To ensure that capillarity and viscous forces dominate the dynamics of the fluids, 
inertia must be small compared to both of these forces.  We enforce this situation by neglecting 
the convective term in Eq.~(\ref{eq:ns1}).   As for compressibility,  we consider low Mach number flows, 
which we achieve by keeping $U \ll c_s$. For our scheme, it suffices to set $U \leq 0.01$.

Our goal is to explore the viscous fingering problem for a wide range in $1/B$. 
Due to computation resource limitations, $\epsilon$ is restricted to $O(10)$ at most
for the majority of runs. To achieve high values of $1/B$, say $O(10^3)$, 
$Ca$ must then be $O(10)$.  Our strategy is to keep the interface velocity and the viscosity 
in ranges of $U=O(10^{-2})$ and $\eta = O(10^{-1})$.  Hence, $Ca$ can be tuned by means of the 
surface tension. 
  
The channel is implemented as follows.  We set a rectangular simulation box of dimensions 
$Nx\times Ny\times Nz.$  Due to the flow symmetry, we simulate only one fourth of the real 
channel by setting boundary conditions as follows: $\partial_y \rho v_y (x,y=0,z)=\partial_y \rho v_y (x,y=W/2,z) =0$, 
$\partial_y \phi v_y (x,y=0,z) = \partial_y \phi v_y (x,y=W/2,z) 0 $, $\partial_z \rho v_z (x,y,z=0) = \partial_z \phi v_z (x,y,z=0) = 0.$

\subsection{Linear Stability in the two-dimensional limit}
\label{subsec:LAS}
\begin{figure}
\includegraphics[width=0.45\textwidth]{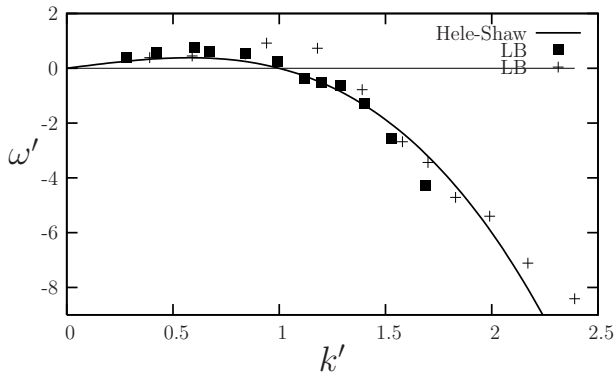}
\caption{Dispersion relation for the linear stability of the interface. Simulation parameters (in simulation units) are 
$\sigma=0.046$, $\eta=0.1$, $b=11.0$ for all runs; $(+)$ $\Delta \rho g = 3.3 \times 10^{-6}$ and $(\blacksquare)$ $\Delta \rho g = 6.6 \times 10^{-6}$\label{fig:dispersion_relation}}
 \end{figure}
We first verify the linear stability of the interface, \textit{i.e.}, the behavior of an initially flat interface 
that has been subjected to a small perturbation.  We study fluids of equal viscosities, so the instability is
gravitationally driven.  This is done by fixing the body force term in Eq.~(\ref{eq:ns1}) 
as $\rho g = 8 \eta /b^2 U_{exp} (\phi+1)/2$, where $U_{exp}$ is the maximum expected velocity 
for a Poiseuille flow.   The modified capillary number reduces to $1/B= W^2\Delta{\rho g }/ \sigma$, 
In this case, the linear stability analysis of the interface evolution of the averaged equations yields the 
dispersion relation
\begin{equation}
\omega(k)=\frac{b^2}{24\eta}k(\Delta \rho g - \sigma k^2),
\label{eq:disp_rel}
\end{equation}
where $\omega$ is the exponential growth rate of a sinusoidal perturbation 
to the flat interface solution. The perturbation is characterized by its wavelength, 
$\Lambda=2\pi/k$. By considering dimensionless frequencies $\omega^\prime = \omega / (U/2W) B^{1/2}$ and wavenumbers, $k^\prime=WB^{1/2}k$, the dispersion relation becomes universal, \textit{i.e.},
$\omega^\prime(k^\prime)=k^\prime(1-{k^\prime}^2).$
   
We prepare a base flow corresponding to a flat interface in the $xy$ plane that propagates 
at constant velocity.  The interface in the $xz$ plane is nearly flat throughout the simulation, 
so the system is effectively two-dimensional. Once the base flow is fully developed, the interface is shifted according to a 
single mode perturbation of wavelength $\Lambda=W$ and an initial small amplitude.
We follow the evolution of the amplitude, $A(t)$, which is measured as 
$A(t)=x_{tip}(t)-\bar x(t),$ where $x_{tip}$ and $\bar x(t)$ are the interface tip and mean interface 
positions respectively.  The growth rate, $\omega$, is extracted as a linear fit of $\log(A(t))$ \emph{vs} $t$. 

Fig.~{\ref{fig:dispersion_relation}} shows a comparison between the universal dispersion relation and our results.  
To quantify the degree of accuracy of these results, we fit our data to the general form $ak^\prime(b-c{k^\prime}^2)$. 
We find a most unstable mode at $k^\prime_{max}\simeq 0.56$ and a first unstable mode at 
$k^\prime_0\simeq 0.96$, both 4\% below the exact result.    

\subsection{Viscous Fingers}       
\label{subsec:VF}
In a preceding study\cite{Ledesma01}, we have shown that it is possible to control the generation of a thin film 
in the channel by adjusting the diffusivity of the order parameter.  Although for usual 
experimental conditions this is not a relevant parameter (it might be relevant in nano-channels), 
it gives the possibility of elucidating the role of a thin film in viscous fingers.   
Diffusivity is accounted for by a P\'eclet number, $Pe=Ub/D$, where $D$ is the diffusion coefficient.   
By combining the effects of $Pe$ and $Ca$, one can either suppress or induce the formation of a 
thin film.  In particular, a small value of the product $CaPe$ results in the suppression of thin films, 
while the contrary is obtained for high $CaPe$.  Results from the preceding work give a penetration threshold 
of $CaPe \simeq 10^{-1}$.  

The strategy is to study first fingers for which $CaPe \leq 10^{-1}$ and then extend this simulations 
to $CaPe \gg 10^{-1}$. 
 
\subsubsection{Meniscus Regime}
\label{subsubsec:VFM}
\begin{table}[b!]
\caption{Control parameters and finger width for runs in of the meniscus regime.\label{tab:results2d}}
\begin{ruledtabular}
\begin{tabular}{lccccc}
Run & $\epsilon$  &   $Ca$ & $CaPe$ & $1/B$ &$\lambda$\cr\hline
(a)&0.17  &  0.11  &  0.08  & 99   &  0.709   \cr
(b)&0.17  &  0.22  &  0.16  & 198  &  0.675   \cr
(c)&0.17  &  0.45  &  0.19  & 290  &  0.640   \cr
(d)&0.06  &  0.11  &  0.04  & 522  &  0.558   \cr
(e)&0.06  &  0.19  &  0.02  & 1045 &  0.525   \cr
(f)&0.06  &  0.23  &  0.03  & 1672 &  0.523   \cr
(g)&0.06  &  0.48  &  0.11  & 2090 &  0.529   \cr
(h)&0.06  &  0.68  &  0.21  & 3136 &  0.518   \cr
(i)&0.04  &  0.74  &  0.26  & 4175 &  0.521   \cr
(j)&0.05  &  0.76  &  0.27  & 6012 &  0.519   \cr
\end{tabular}
\end{ruledtabular}
\end{table}
\label{sec:2D}

\begin{figure*}[t!]
\includegraphics[width=0.9\textwidth]{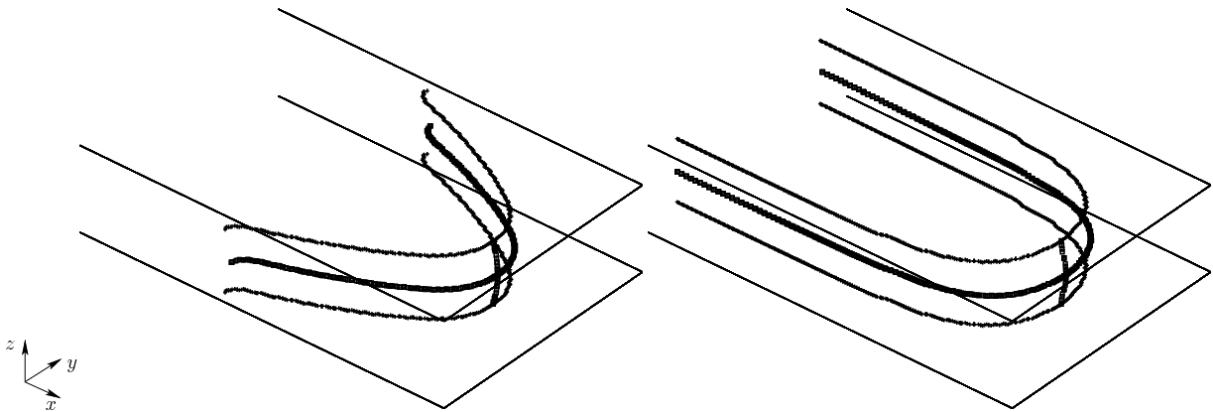}
\caption{Interface snapshots at two different times for $\epsilon=0.17$(the plot is off-scale), $1/B=99$ and $CaPe=0.08$. 
The thick line parallel to the $xy$ plane corresponds to the leading interface, while the thick line
parallel to the $xz$ plane corresponds to the interface projection in the center of the channel. 
Thin lines correspond to the contact lines. The first snapshot corresponds to $t=0.11 b/U$, while the second snapshot, at $t=17.74b/U$, 
corresponds to the steady state finger.
\label{fig:st}}
\end{figure*}

\begin{figure}
\centering
\includegraphics[width=0.45\textwidth]{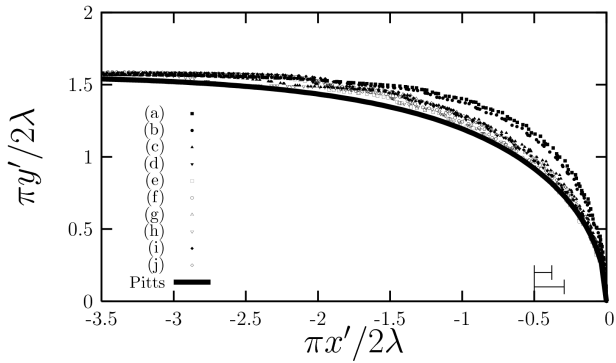}
\caption{Collapsed interface profiles in the $xy$ plane for the meniscus regime. Parameter values corresponding to each symbol can be consulted in Table~\ref{tab:results2d}.  The error (small bar at the right bottom) corresponds to one lattice spacing. The larger bar indicates the size of the diffuse 
interface, approximately $3\xi$.\label{fig:shapes.2d}}
\end{figure}

We first study fingers for which no film of displaced fluid develops in the $xz$ plane of the channel.  
We carry out simulations with modified capillary numbers in the range $100\leq 1/B \leq 6000$.
We have studied different geometries, ranging from $\epsilon=0.17$ to $\epsilon= 0.04$. The aspect ratio is decreased
by decreasing the channel thickness. We summarize the simulation parameters in Table~\ref{tab:results2d}.

For each run we observe the usual phenomenology for the leading interface.      
During the early stages of interface evolution, the amplitude of the perturbation grows until 
a finger emerges and widens.  This stage is followed by a relaxation of the interface shape, until a 
Saffman-Taylor finger develops.  The finger propagates with a steady velocity $U$, leaving behind 
a growing region where the finger has flat sides. In this region a constant finger width, $\lambda W$, can be defined.  
As for the channel thickness, we observe that the initially flat interface rapidly relaxes to a meniscus, which also has 
a steady shape.  In Fig.~(\ref{fig:st}) we show a three-dimensional plot of the interface for run (a) in Table~\ref{tab:results2d}  
at two different times.  In the plot we show both the contact lines and the leading interface; both contact lines 
follow the leading interface. 

To check for consistency in the steady state solution we use the semiempirical interface profile 
obtained by Pitts\cite{Pitts01}, which reproduces experimental results accurately for a wide range of
finger widths.  The equation for the interface shape reads, 
\begin{equation}
\cos(\pi y^\prime /2\lambda )=\exp(\pi x^\prime/2\lambda).
\label{eq:Pitts}
\end{equation}
where $x^\prime$ and $y^\prime$ measure the distance from the finger tip in units of half 
the channel width.  The natural scalings in this equation are $\pi x^\prime /2\lambda$ and 
$\pi y^\prime /2\lambda$.   Consequently, all profiles should collapse into a single curve if these scalings 
are used.  Fig.~\ref{fig:shapes.2d} shows plot of interface profiles 
corresponding to runs of Table~\ref{tab:results2d}.   As expected, all interface profiles fall in the
same universal curve within error.  In addition, our collapse is in fair agreement with Eq.(\ref{eq:Pitts}).

\begin{figure}
\begin{centering}
\includegraphics[width=0.45\textwidth]{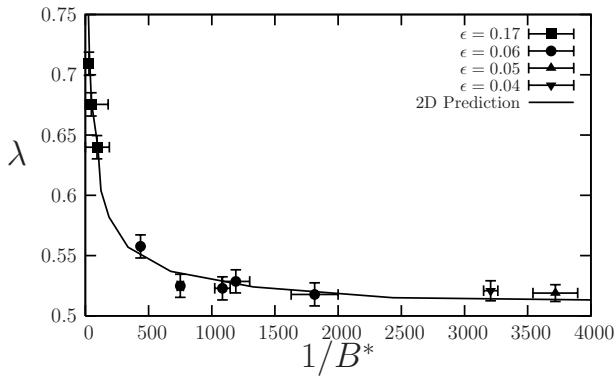}
\end{centering}
\caption{Finger width as a function of the rescaled control parameter $1/B^*$ in the meniscus regime. \label{fig:rescaledresults}}
\end{figure}

The selection rule in the viscous fingering problem is expressed 
as the functional dependence of the finger width with the modified control
parameter. We compare our results with the $\lambda$ \emph{vs.} $1/B$ prediction of McLean and Saffman.  
We find that runs with $\epsilon=0.17$ show wider fingers than predicted, while runs with smaller $\epsilon$ agree better 
with the two-dimensional result.  Even in the absence of a thin film, the $xz$ interface projection has a certain curvature. 
This can be accounted for by defining an effective surface tension in terms of the radii of curvature of the interface, which 
we are able to measure directly. The effective surface tension then reads $\sigma^* = \sigma(1+R_\parallel/R_\perp)$.  
The correction factor in this expression is given by the quantity in parentheses, 
which increases for strongly curved meniscus.  The rescaled control parameter then reads $1/B^*=(1/B)/(1+R_\parallel/R_\perp).$  
Of course this correction should be more evident in the low $1/B$ region, where $\lambda$ varies rapidly with the modified control 
parameter.  In Fig.~\ref{fig:rescaledresults}  we show a plot of $\lambda$ \emph{vs.} $1/B^*$.  Points fall 
on the McLean-Saffman curve for the wide range of $1/B^*$ considered, regardless of the aspect ratio.  

\subsubsection{Thin Film Regime}
\label{subsubsec:VFTF}
\begin{table}[b!]
\caption{Control parameters and finger width for thin film regime runs.\label{tab:results3d}}
\begin{ruledtabular}
\begin{tabular}{lccccc}
Run &$\epsilon$  &   $Ca$ & $CaPe$ & $1/B$ &$\lambda$\cr\hline
(a)&0.25  &  2.80  &  12.32  & 835  &  0.592   \cr
(b)&0.25  &  3.36  &  17.74  & 1002 &  0.589   \cr
(c)&0.25  &  6.61  &  68.61  & 2004 &  0.569   \cr
(d)&0.25  &  15.9  &  400.41 & 4003 &  0.558   \cr
(e)&0.35  &  8.96  &  1515  & 1403 &  0.549   \cr
(f)&0.49  &  34.7  &  4330  & 5247 &  0.527   \cr
(g)&0.64  &  50.9  &  3973  & 5247 &  0.517   \cr
(h)&0.78  &  68.5  &  7192  & 5247 &  0.508   \cr
(i)&1.00  &  91.9  &  8019  & 5247 &  0.493   \cr
(j)&1.00  &  131   &  156598& 5430 &  0.494   \cr
\end{tabular}
\end{ruledtabular}
\end{table}

\begin{figure}[!t]
\centering
\includegraphics[width=0.45\textwidth]{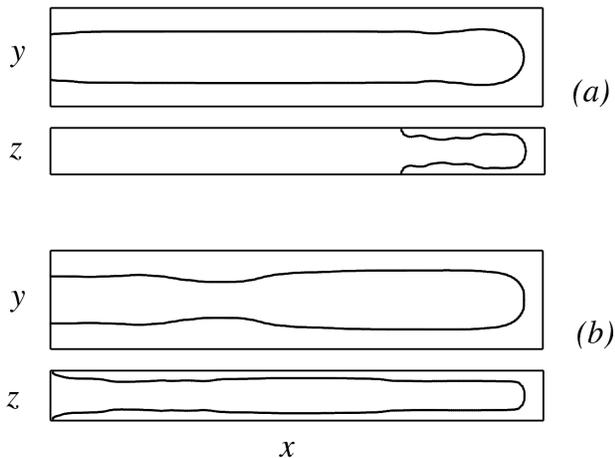}
\caption{Interface projections in the $xy$ and $xz$ planes for runs with different $CaPe$ values.  Plots correspond 
to the same simulation time.  (a) $CaPe=0.85$, (b) $CaPe=4.44$. \label{fig:test156.157.159}}
\end{figure}
We now extend our simulations to fingers in the film regime.  Penetration in the $xz$ plane occurs for high $CaPe$, 
so we choose to sample $1/B$ at fixed $D$. Consequently,  $CaPe$ increases with increasing $1/B$.  
To resolve the thin film correctly we must take into account the finite size of the interface. 
As explained in Ref.\cite{Ledesma01}, the thin film is insensitive to the channel thickness 
already for $b = 23$.  We therefore choose sufficiently thick channels.  
We explore a wide range of aspect ratios, $0.25 \leq \epsilon \leq 1.0 $ 
and modified capillary numbers, $800 \leq 1/B \leq 5300$. 

We first explore the $CaPe\sim O(1)$ region, close to the penetration threshold.  
In Fig.~\ref{fig:test156.157.159} we show interface projections in the $xy$ and $xz$ planes 
located at $z=b/2$ and $y=W/2$ respectively.  We show two sets of interfaces, 
corresponding to two different $CaPe$ values; (a)$CaPe=0.85$ and (b)$CaPe=4.44$.  In Fig.~\ref{fig:test156.157.159}(a) the interface in the $xz$ 
plane presents a penetrating structure,  but a well developed film is absent.  The finger in the $xy$ plane 
is not well developed either, and it presents an anomalous tip.  Conversely, in Fig.~\ref{fig:test156.157.159}(b) 
both interface projections describe well developed fingers.  It is then clear that deviations from the Saffman-Taylor finger 
in the $xy$ plane are correlated to the interface structure in the $xz$ plane.   An interesting feature of the run corresponding 
to Fig.~\ref{fig:test156.157.159}(a) is that the 
the $xz$ interface structure is persistent.  This means that the length of the finger
in the $xz$ plane is constant in time, a consequence of the slip velocity of the contact line.  
The diffusion strength is not large enough to maintain a meniscus, which  in the one hand makes the slip velocity 
smaller than the channel velocity. Nevertheless, as the interface relaxes to a thin film shape, curvature deviations 
from equilibrium increase the slip velocity, making the contact line advance to restore the meniscus shape. 

\begin{figure*}
\includegraphics[width=0.9\textwidth]{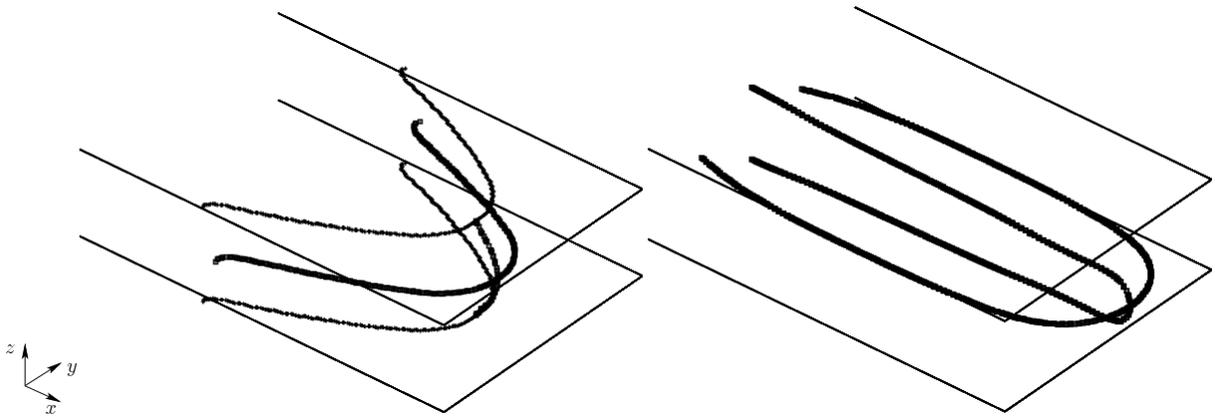}
\caption{Interface snapshots at two different times for $\epsilon=0.25$(the plot is off-scale), $1/B=1002$ and $CaPe=17.74$.  
Thick lines correspond to the $xy$ and $xz$ interface projections in the center of the channel.  Thin lines correspond to the contact lines.
Times are $t=0.57b/U$ and $t=28.84b/U$. 
\label{fig:test160}}
\end{figure*}
\begin{figure}
\centering
\includegraphics[width=0.45\textwidth]{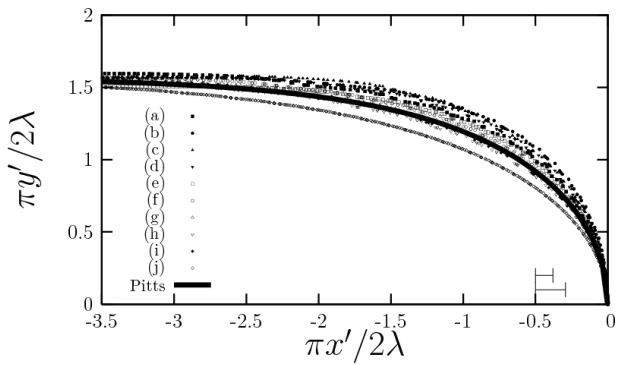}
\caption{Rescaled interface profiles for the thin film regime. 
Symbols correspond to data presented in Table~\ref{tab:results3d}.  
The bars in the bottom at the right indicate the error bar and diffuse interface size as in Fig.~\ref{fig:shapes.2d}.\label{fig:shapes.3d}}
\end{figure}

We next explore the range $CaPe \geq O(10)$ for which simulation parameters and observed finger widths are summarized 
in Table~\ref{tab:results3d}.  In Fig.~\ref{fig:test160} we present snapshots of the three dimensional interface 
at two different times for run (b) in Table~\ref{tab:results3d}.  
The first snapshot corresponds to the early stage of the finger formation.  
Looking at the interface projections in the $xy$ plane, we see that the contact line(light line) is close 
to the leading interface(dark line) and no film is present in the $xz$ plane.  In the next snapshot the contact line has moved away from the tip, thus giving rise to the growth of a wetting film. The shape of the finger is
in agreement with the typical morphology found in experiments.  To illustrate this, in Fig.~\ref{fig:shapes.3d} 
we compare the shape of the finger to Eq.~(\ref{eq:Pitts}).  Within error, our profiles are consistent 
with Pitts shape.  
  
\begin{figure}[b!]
\begin{centering}
\includegraphics[width=0.45\textwidth]{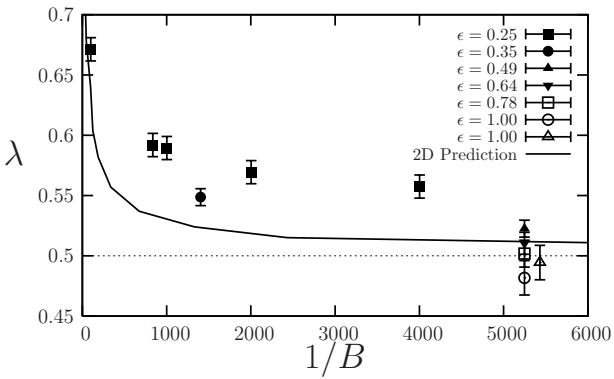}
\end{centering}
\caption{Finger width as a function of $1/B$. \label{fig:results3D}}
\end{figure}

Fig.~\ref{fig:results3D} shows the measured finger width as a function of $1/B$.  The lowest aspect 
ratio we have considered corresponds to $\epsilon=0.25$(runs (a)-(d) in Table~\ref{tab:results3d}).  
We see that for all $1/B$ values considered the finger width falls above the McLean-Saffman prediction. 
We increase the aspect ratio to $\epsilon=0.35$(run (e) in Table~\ref{tab:results3d}). As a result, the 
measured finger width decreases. Runs for which $\epsilon$ is larger confirm this tendency in a systematic way.   
Tests (f)-(j) in the same table correspond to a fixed value of $1/B$ with increasing $\epsilon$. We find that for 
sufficiently large $\epsilon$ the finger width goes below the one-half theoretical limit of McLean and Saffman. 
 
\section{Discussion}
\label{sec:Disc} 
\begin{figure}[t!]
\begin{centering}
\includegraphics[width=0.45\textwidth]{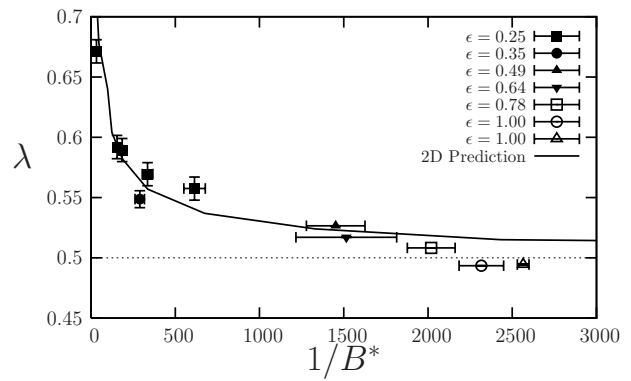}
\end{centering}
\caption{Finger width as a function of the rescaled control parameter for the thin film regime. \label{fig:rescaledresultsthinfilm}}
\end{figure}
Our results show that the finger width decreases with increasing aspect ratio.  To maintain $1/B$ fixed while varying the aspect ratio of the channel, one has to vary $Ca$ accordingly.  As a consequence, the film thickness and the capillary pressure 
are altered.  If we  increase the aspect ratio(as in the high-$1/B$ region in Fig.~\ref{fig:results3D}), then $Ca$ 
must decrease to keep $1/B$ fixed.  As a consequence, the film thickness and the capillary pressure 
decrease as $\epsilon$ increases, which is consistent with a narrower finger. 

This behavior has been observed, for instance, in experiments by Tabeling, Zocchi and Libchaber\cite{libchaber02}, 
and addressed in numerical calculations by Reinelt\cite{reinelt} where the effect of the thin film was introduced 
perturbatively in the two-dimensional model.  Experiments suggest that, for high $1/B$, increasing the cell aspect 
ratio has a thinning effect on the finger, which is what we observe in our simulations.  Results of Reinelt 
suggest the opposite tendency. 
 
The aforementioned experiments were done at small $\epsilon$ and $Ca$, and at high viscosity contrast, defined as 
$c=(\eta_2-\eta_1)/(\eta_2+\eta_1)$.  As we have shown in Ref.\cite{Ledesma01}, the thin film 
thickens as $c\rightarrow 1$.  Under these conditions, experiments show that the finger width goes below the one-half limit 
even for cells with $\epsilon=0.009$.  This is due to the small thickness of the film which is a consequence of the low $Ca$ 
and high $c$ values used in experiments.  In our simulations the thin film is about $t/b\simeq 0.25$, much thicker than the experimental 
value, $t/b\simeq 0.05$.  As a consequence, curvature effects in our simulations are strong enough to keep the finger width above one 
half even for high values of $\epsilon$.  To achieve the experimental regime
thinner film should be considered. We have considered a cell aspect ratio of $\epsilon \simeq 0.05$ and $c=0.9$.  
Nevertheless, $Ca$ is still too large, the film is then thick enough to keep us in the low $1/B^*$ regime, where 
the finger width is still larger than one half of the channel width.  Due to computational limitations we do not explore 
smaller $\epsilon$.

The fact that for a given $1/B$ there exist different finger widths for different aspect ratios
raises the doubt of $1/B$ as being the only control parameter present in the system. To this end, we 
compute the rescaled surface tension $\sigma^*=\sigma\left(1+R_\parallel/R_\perp\right)$, 
where the radii of curvature are measured at the finger tip. We then rescale the control parameter 
according to $1/B^*=(\sigma/\sigma^*)1/B$.  In Fig.~\ref{fig:rescaledresultsthinfilm} we show a plot of the finger width as 
a function of the rescaled control parameter.  At low $1/B^*$, results agree with McLean-Saffman results.  
We conclude that in this region the finger is effectively two dimensional and that three-dimensional effects 
can be accounted for even at $Ca \sim 1$.
 
At high values of the rescaled control parameter, our results deviate systematically from the McLean-Saffman curve, 
until the finger width goes below the one half limit of the two-dimensional theory.  This behavior is qualitatively 
different from the one found for the meniscus regime, in which the McLean-Saffman curve could be recovered at any 
value of $1/B^*$. Hence, we conclude that deviations from two-dimensionality are caused by the thin film. 

An important feature in the $\lambda$ \emph{vs.} $1/B^*$ plot is that finger width appears 
to be determined by $1/B^*$ uniquely. This suggests 
that $1/B^*$ is the only control parameter of the problem.

We have explored a region of values of the aspect ratio between the Saffman-Taylor($\epsilon \rightarrow 0$)
and Rayleigh-Taylor($\epsilon = 1$) limits of the fingering instability.  In both limits, the relevant control 
parameter appears to be an effective modified capillary number.  In addition, the interface shape is remarkably 
universal, as suggested by Figs.\ref{fig:shapes.2d} and~\ref{fig:shapes.3d}. 
 
\section{Conclusions}
\label{sec:Conclusions}
We have carried out a detailed study of the viscous fingering problem in three-dimensional
channels for fluids of different densities and viscosities.  We have studied the single finger solution 
for systems in which either a thin film develops across the channel thickness or a meniscus is stabilized. 

For systems in which no thin film is present, McLean-Saffman two-dimensional results describe the dependency 
of the finger width as a function of a rescaled modified capillary number, $1/B^*$, which has a 
correction that depends curvature of the interface direction of the channel thickness. This holds for arbitrary high values of $1/B^*$, 
evidencing that a complete displacement across the channel thickness renders the problem two-dimensional.  

We have extended our studies to situations where a thin film develops across the channel.  
We find different values of the finger width when changing the channel aspect ratios at fixed modified 
capillary number, an observation that is consistent with previous experiments\cite{libchaber02}.  This non-uniqueness 
seems to disappear as the control parameter is corrected by curvature effects associated to the thin film, \textit{i.e.}, 
when the finger width is compared to $1/B^*$.  

For low $1/B^*$, the finger width collapses to the McLean-Saffman curve.   However, at high $1/B^*$ 
the finger width deviates from this curve, and goes bellow the limit of one half in units of the channel 
width.

Our work is done at high values of the capillary number.  Consequently, the effective capillary pressure 
in our simulations is large enough to keep the finger above the one-half limit of the two-dimensional 
theory for high values of the channel aspect ratio. Experiments in Refs.\cite{libchaber,libchaber02} were done in cells 
with $\epsilon \simeq 0.03$ and at $Ca \simeq 10^{-3}$,  a regime that falls beyond the scope of this 
work for computational reasons.  Nonetheless, for low $1/B^*$, we recover the same results, 
indicating that the same mechanisms hold, even if the actual aspect ratio and capillary number
are not the same in experiments and simulations.

To our knowledge, experiments of fingering in high aspect ratio channels have not been conducted so far.  
Our results could be confirmed, for instance, in micro-channels, where the aspect ratio is typically large and in which 
the meniscus to thin film transition could be observed.  This is then an open question for experimentalists
to confirm. 

\begin{acknowledgments}
We acknowledge financial support from Direcci\'on General de Investigaci\'on (Spain) under projects FIS\ 2006-12253-C06-05 and FIS\ 2005-01299.
R.L.-A. wishes to acknowledge support from CONACyT (M\'exico) and Fundaci\'on Carolina(Spain).
Part of the computational work herein was carried on in the MareNostrum Supercomputer at Barcelona Supercomputing Center.
\end{acknowledgments}


\end{document}